\documentclass[%
	conference,
]{IEEEtran}

\usepackage[utf8]{inputenc}
\usepackage{xspace}

\usepackage{standalone}		%
\usepackage{xstring}

\usepackage[binary-units=true]{siunitx}
\usepackage{graphicx}

\usepackage{tikz}
\tikzset{font=\sffamily\footnotesize}

\usetikzlibrary{spy,calc,shapes.arrows,patterns}
\usepackage{pgfplots}
\pgfplotsset{compat=newest}
\pgfplotsset{
    box plot/.style={
        /pgfplots/.cd,
        fill=blue!10,
        only marks,
        mark=-,
        mark size=0.2em,
        /pgfplots/error bars/.cd,
        y dir=plus,
        y explicit,
    },
    box plot box/.style={
        /pgfplots/error bars/draw error bar/.code 2 args={%
            \draw  ##1 -- ++(.2em,0pt) |- ##2 -- ++(-.2em,0pt) |- ##1 -- cycle;
        },
        /pgfplots/table/.cd,
        y index=2,
        y error expr={\thisrowno{3}-\thisrowno{2}},
        /pgfplots/box plot
    },
    box plot top whisker/.style={
        /pgfplots/error bars/draw error bar/.code 2 args={%
            \pgfkeysgetvalue{/pgfplots/error bars/error mark}%
            {\pgfplotserrorbarsmark}%
            \pgfkeysgetvalue{/pgfplots/error bars/error mark options}%
            {\pgfplotserrorbarsmarkopts}%
            \path ##1 -- ##2;
        },
        /pgfplots/table/.cd,
        y index=4,
        y error expr={\thisrowno{2}-\thisrowno{4}},
        /pgfplots/box plot
    },
    box plot bottom whisker/.style={
        /pgfplots/error bars/draw error bar/.code 2 args={%
            \pgfkeysgetvalue{/pgfplots/error bars/error mark}%
            {\pgfplotserrorbarsmark}%
            \pgfkeysgetvalue{/pgfplots/error bars/error mark options}%
            {\pgfplotserrorbarsmarkopts}%
            \path ##1 -- ##2;
        },
        /pgfplots/table/.cd,
        y index=5,
        y error expr={\thisrowno{3}-\thisrowno{5}},
        /pgfplots/box plot
    },
    box plot median/.style={
        /pgfplots/box plot
    },
    boxplot/every median/.style={
    	ultra thick,dashed,cyan
    }
}

\usepackage{todonotes}
\ifCLASSINFOpdf
\else
\fi
\usepackage{amsmath}

\usepackage{sansmath}	%

\usepackage[%
IEEE,%
]{copyright-overlay}
\CopyrightAuthor{Robert Falkenberg}
\CopyrightYear{2016}
\ConferenceName{2016 IEEE 84th Vehicular Technology Conference (VTC-Fall)}
\Doi{10.1109/VTCFall.2016.7880932}
\Bibtex{%
@InProceedings\{Falkenberg/etal/2016a,\textCR\textLF
\textHT{}author         = \{Robert Falkenberg and Christoph Ide and Christian Wietfeld\},\textCR\textLF
\textHT{}title          = \{Client-Based Control Channel Analysis for Connectivity Estimation in \{LTE\} Networks\},\textCR\textLF
\textHT{}booktitle      = \{2016 IEEE 84th Vehicular Technology Conference (VTC-Fall)\},\textCR\textLF
\textHT{}year           = \{2016\},\textCR\textLF
\textHT{}address        = \{Montreal, Canada\},\textCR\textLF
\textHT{}month          = \{Sep\},\textCR\textLF
\textHT{}organization   = \{IEEE\},\textCR\textLF
\textHT{}doi            = \{10.1109/VTCFall.2016.7880932\}\textCR\textLF
\}
}

\usepackage[]{acronym}

\acrodef{RSSI}[\ensuremath{\mathrm{RSSI}}]{Received Signal Strength Indicator}
\acrodef{RSRP}[\ensuremath{\mathrm{RSRP}}]{Reference Signal Received Power}
\acrodef{RSRQ}[\ensuremath{\mathrm{RSRQ}}]{Reference Signal Received Quality}
\acrodef{SINR}[\ensuremath{\mathrm{SINR}}]{Signal to Interference and Noise Ratio}
\acrodef{PRB}[\text{PRB}]{Physical Resource Block}
\acrodef{RB}[\text{RB}]{Resource Block}
\acrodef{CRC}[CRC]{Cyclic Redundancy Check}
\acrodef{DCI}[DCI]{Downlink Control Information}
\acrodef{RNTI}[RNTI]{Radio Network Temporary Identifier}
\acrodef{SI-RNTI}[\ensuremath{\mathrm{SI-RNTI}}]{System-Information \acs{RNTI}}
\acrodef{CCE}[CCE]{Control Channel Element}
\acrodef{PDCCH}[PDCCH]{Physical Downlink Control Channel}
\acrodef{UE}[UE]{User Equipment}
\acrodef{eNodeB}[eNodeB]{evolved NodeB}
\acrodef{SDR}[SDR]{Software Defined Radio}
\acrodef{STG}[STG]{Smart Traffic Generator}
\acrodef{DUT}[DUT]{Device Under Test}
\acrodef{OAI}[OAI]{Open Air Interface}
\acrodef{MCS}[MCS]{Modulation and Coding Scheme}
\acrodef{OFDM}[OFDM]{Orthogonal Frequency Division Multiplexing}
\acrodef{DL}[DL]{Downlink}
\acrodef{UL}[UL]{Uplink}
\acrodef{C3ACE}[C$^3$ACE]{Client-based Control Channel Analysis for Connectivity Estimation}
\acrodef{TCP}[TCP]{Transmission Control Protocol}

\newcommand{\RSSI}{\ac{RSSI}\xspace}
\newcommand{\RSRP}{\ac{RSRP}\xspace}
\newcommand{\RSRQ}{\ac{RSRQ}\xspace}

\newcommand{\RB}{\ac{RB}\xspace}
\newcommand{\RBs}{\acp{RB}\xspace}

\newcommand{\CRC}{\ac{CRC}\xspace}
\newcommand{\DCI}{\ac{DCI}\xspace}
\newcommand{\RNTI}{\ac{RNTI}\xspace}
\newcommand{\RNTIs}{\acp{RNTI}\xspace}

\newcommand{\CCE}{\ac{CCE}\xspace}
\newcommand{\CCEs}{\acp{CCE}\xspace}
\newcommand{\PDCCH}{\ac{PDCCH}\xspace}
\newcommand{\UE}{\ac{UE}\xspace}
\newcommand{\UEs}{\acp{UE}\xspace}
\newcommand{\eNB}{\ac{eNodeB}\xspace}

\newcommand{\SDR}{\ac{SDR}\xspace}

\newcommand{\STGs}{\acp{STG}\xspace}
\newcommand{\DUT}{\ac{DUT}\xspace}
\newcommand{\OAI}{\ac{OAI}\xspace}
\newcommand{\MCS}{\ac{MCS}\xspace}

\newcommand{\cccace}{\ac{C3ACE}\xspace}
\newcommand{\TCP}{\ac{TCP}\xspace}

\newcommand{\dBm}{dBm}

\newcommand{\hex}[1]{\texttt{0x#1}\xspace}

\newcommand{\iperf}{\textsc{iPerf3}\xspace}

\renewcommand{\baselinestretch}{0.99}
\begin{document}
\title{Client-Based Control Channel Analysis for Connectivity Estimation in LTE Networks}

\author{\IEEEauthorblockN{Robert Falkenberg, Christoph Ide and Christian Wietfeld}
\IEEEauthorblockA{Communication Networks Institute\\TU Dortmund University\\44227 Dortmund, Germany\\
Email: \{Robert.Falkenberg, Christoph.Ide, Christian.Wietfeld\}@tu-dortmund.de}%
}

\maketitle

\begin{abstract}
Advanced Cyber-Physical Systems aim for the balancing of restricted local resources of deeply embedded systems with cloud-based resources depending on the availability of network connectivity:
in case of excellent connectivity, the off-loading of large amounts of data can be more efficient than the local processing on a resource-constraint platform, while this latter solution is preferred in case of limited connectivity.
This paper proposes a Client-Based Control Channel Analysis for Connectivity Estimation (C$^3$ACE), a new passive probing mechanism to enable the client-side to estimate the connection quality of 4G networks in range.
The results show that by observing and analyzing the control traffic in real-time, the number of active user equipment in a cell can be determined with surprising accuracy (with errors well below $10^{-6}$).
The specific challenge addressed in this paper lies in a dedicated filtering and validation of the DCI (Downlink Control Information).
In a subsequent step the data rates to be expected can be estimated in order to enable decision about the choice of network and the timing of the data offloading to the cloud.
The proposed methods have been implemented and validated leveraging the SDR OpenAirInterface, a real-life LTE network and a distributed load generator producing a scalable network traffic by a number of LTE User Equipment.
\end{abstract}

\IEEEpeerreviewmaketitle

\PrintCopyrightOverlay

\section{Introduction}

Forecasting the expected data rate of an LTE connection might be useful in many cases.
For example, resource constrained Cyber-Physical Systems could decide whether it is worthwhile to offload certain computations to a cloud service or to process them locally, if offloading would require more energy or time.
Moreover, this information would enable vehicles to select the currently fastest link for transmitting telemetry data or emergency messages, which might not always be the mobile network but a locally available WiFi hotspot or a Car-to-Car link.
Especially warnings about an upcoming hidden tail of a motorway traffic jam might be delayed by the high load of the local LTE cells as a consequence of the high density of devices in this place.

Current LTE networks do not provide any passive indicators to reliably estimate the data rate for an ongoing transmission.
Commercial off-the-shelf  \UE can only obtain a data rate retrospectively after an actual transmission.
Existing indicators like \RSSI, \RSRP and \RSRQ allow for a rating of the current radio conditions and therefore an estimation of the highest achievable data rate.
But this estimation suffers of large uncertainties as the true data rate highly depends on the current load of the associated \eNB.
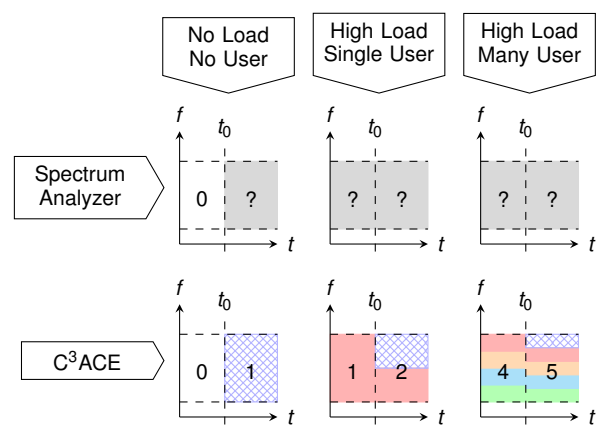
\begin{figure}[b!]  	
\centering		  
	\begin{tikzpicture}
\begin{sansmath}

\def\plotdistanceX{2cm}
\def\plotdistanceY{2.3cm}

\def\lenX{1.3cm}
\def\lenY{1.5cm}

\def\flow{2mm}
\def\fhigh{1.1cm}

\def\txhigh{1.3cm}

\def\ttx{6mm}
\def\ttylow{-1mm}
\def\ttyhigh{1.3cm}

\def\colorA{blue!30}
\def\colorB{red!30}
\def\colorC{orange!30}
\def\colorD{cyan!30}
\def\colorE{green!30}
\def\colorBlind{gray!30}

\tikzstyle{capt} = [node distance=0.5cm and 0.2cm,single arrow,draw,single arrow tip angle=150,single arrow head extend=0cm]
\tikzstyle{captll} = [node distance=0.5cm and 0.2cm,single arrow,draw,single arrow tip angle=120,single arrow head extend=0cm]

\coordinate (org) at (0,0);

\draw[fill,\colorA,pattern=crosshatch,pattern color=\colorA] ([xshift=\ttx, yshift=\flow]org) rectangle ([xshift=\txhigh,yshift=\fhigh]org);

\draw[-stealth] (org) to +(\lenX,0) node[right] {$t$};
\draw[-stealth] (org) to +(0,\lenY) node[above] {$f$};

\draw[dashed] ([yshift=\flow]org) to +(\txhigh,0);
\draw[dashed] ([yshift=\fhigh]org) to +(\txhigh,0);

\draw[dashed] ([xshift=\ttx, yshift=\ttylow]org) to ([xshift=\ttx, yshift=\ttyhigh]org)  node[above] {$t_0$} ;

\node[] at ([xshift=0.5*\ttx,yshift=\ttylow+0.5*(\ttyhigh-\ttylow)]org) {0};
\node[] at ([xshift=\ttx+0.5*(\txhigh-\ttx),yshift=\ttylow+0.5*(\ttyhigh-\ttylow)]org) {1};

\coordinate (ll1) at ([yshift=0.5*\lenY]org);

\coordinate (org) at (\plotdistanceX,0);

\draw[fill,\colorB] ([yshift=\flow]org) rectangle ([xshift=\ttx,yshift=\fhigh]org);

\draw[fill,\colorB] ([xshift=\ttx, yshift=\flow]org) rectangle ([xshift=\txhigh,yshift=\flow+0.5*(\fhigh-\flow)]org);
\draw[fill,\colorA,pattern=crosshatch,pattern color=\colorA] ([xshift=\ttx, yshift=\flow+1*0.5*(\fhigh-\flow)]org) rectangle ([xshift=\txhigh,yshift=\flow+2*0.5*(\fhigh-\flow)]org);

\draw[-stealth] (org) to +(\lenX,0) node[right] {$t$};
\draw[-stealth] (org) to +(0,\lenY) node[above] {$f$};

\draw[dashed] ([yshift=\flow]org) to +(\txhigh,0);
\draw[dashed] ([yshift=\fhigh]org) to +(\txhigh,0);

\draw[dashed] ([xshift=\ttx, yshift=\ttylow]org) to ([xshift=\ttx, yshift=\ttyhigh]org)  node[above] {$t_0$} ;

\node[] at ([xshift=0.5*\ttx,yshift=\ttylow+0.5*(\ttyhigh-\ttylow)]org) {1};
\node[] at ([xshift=\ttx+0.5*(\txhigh-\ttx),yshift=\ttylow+0.5*(\ttyhigh-\ttylow)]org) {2};

\coordinate (org) at (2*\plotdistanceX,0);

\draw[fill,\colorE] ([yshift=\flow+0*0.25*(\fhigh-\flow)]org) rectangle ([xshift=\ttx,yshift=\flow+1*0.25*(\fhigh-\flow)]org);
\draw[fill,\colorD] ([yshift=\flow+1*0.25*(\fhigh-\flow)]org) rectangle ([xshift=\ttx,yshift=\flow+2*0.25*(\fhigh-\flow)]org);
\draw[fill,\colorC] ([yshift=\flow+2*0.25*(\fhigh-\flow)]org) rectangle ([xshift=\ttx,yshift=\flow+3*0.25*(\fhigh-\flow)]org);
\draw[fill,\colorB] ([yshift=\flow+3*0.25*(\fhigh-\flow)]org) rectangle ([xshift=\ttx,yshift=\flow+4*0.25*(\fhigh-\flow)]org);

\draw[fill,\colorE] ([xshift=\ttx, yshift=\flow+0*0.20*(\fhigh-\flow)]org) rectangle ([xshift=\txhigh,yshift=\flow+1*0.20*(\fhigh-\flow)]org);
\draw[fill,\colorD] ([xshift=\ttx, yshift=\flow+1*0.20*(\fhigh-\flow)]org) rectangle ([xshift=\txhigh,yshift=\flow+2*0.20*(\fhigh-\flow)]org);
\draw[fill,\colorC] ([xshift=\ttx, yshift=\flow+2*0.20*(\fhigh-\flow)]org) rectangle ([xshift=\txhigh,yshift=\flow+3*0.20*(\fhigh-\flow)]org);
\draw[fill,\colorB] ([xshift=\ttx, yshift=\flow+3*0.20*(\fhigh-\flow)]org) rectangle ([xshift=\txhigh,yshift=\flow+4*0.20*(\fhigh-\flow)]org);
\draw[fill,\colorA,pattern=crosshatch,pattern color=\colorA] ([xshift=\ttx, yshift=\flow+4*0.20*(\fhigh-\flow)]org) rectangle ([xshift=\txhigh,yshift=\flow+5*0.20*(\fhigh-\flow)]org);

\draw[-stealth] (org) to +(\lenX,0) node[right] {$t$};
\draw[-stealth] (org) to +(0,\lenY) node[above] {$f$};

\draw[dashed] ([yshift=\flow]org) to +(\txhigh,0);
\draw[dashed] ([yshift=\fhigh]org) to +(\txhigh,0);

\draw[dashed] ([xshift=\ttx, yshift=\ttylow]org) to ([xshift=\ttx, yshift=\ttyhigh]org)  node[above] {$t_0$} ;

\node[] at ([xshift=0.5*\ttx,yshift=\ttylow+0.5*(\ttyhigh-\ttylow)]org) {4};
\node[] at ([xshift=\ttx+0.5*(\txhigh-\ttx),yshift=\ttylow+0.5*(\ttyhigh-\ttylow)]org) {5};

\coordinate (org) at (0,\plotdistanceY);

\draw[fill,\colorBlind] ([xshift=\ttx, yshift=\flow]org) rectangle ([xshift=\txhigh,yshift=\fhigh]org);

\draw[-stealth] (org) to +(\lenX,0) node[right] {$t$};
\draw[-stealth] (org) to +(0,\lenY) node[above] {$f$};

\draw[dashed] ([yshift=\flow]org) to +(\txhigh,0);
\draw[dashed] ([yshift=\fhigh]org) to +(\txhigh,0);

\draw[dashed] ([xshift=\ttx, yshift=\ttylow]org) to ([xshift=\ttx, yshift=\ttyhigh]org)  node[above] {$t_0$} ;

\node[] at ([xshift=0.5*\ttx,yshift=\ttylow+0.5*(\ttyhigh-\ttylow)]org) {0};
\node[] at ([xshift=\ttx+0.5*(\txhigh-\ttx),yshift=\ttylow+0.5*(\ttyhigh-\ttylow)]org) {?};

\coordinate (l1) at ([xshift=0.5*\lenX, yshift=\lenY]org);
\coordinate (ll2) at ([yshift=0.5*\lenY]org);

\coordinate (org) at (\plotdistanceX,\plotdistanceY);

\draw[fill,\colorBlind] ([yshift=\flow]org) rectangle ([xshift=\txhigh,yshift=\fhigh]org);

\draw[-stealth] (org) to +(\lenX,0) node[right] {$t$};
\draw[-stealth] (org) to +(0,\lenY) node[above] {$f$};

\draw[dashed] ([yshift=\flow]org) to +(\txhigh,0);
\draw[dashed] ([yshift=\fhigh]org) to +(\txhigh,0);

\draw[dashed] ([xshift=\ttx, yshift=\ttylow]org) to ([xshift=\ttx, yshift=\ttyhigh]org)  node[above] {$t_0$} ;

\node[] at ([xshift=0.5*\ttx,yshift=\ttylow+0.5*(\ttyhigh-\ttylow)]org) {?};
\node[] at ([xshift=\ttx+0.5*(\txhigh-\ttx),yshift=\ttylow+0.5*(\ttyhigh-\ttylow)]org) {?};

\coordinate (l2) at ([xshift=0.5*\lenX, yshift=\lenY]org);

\coordinate (org) at (2*\plotdistanceX,\plotdistanceY);

\draw[fill,\colorBlind] ([yshift=\flow]org) rectangle ([xshift=\txhigh,yshift=\fhigh]org);

\draw[-stealth] (org) to +(\lenX,0) node[right] {$t$};
\draw[-stealth] (org) to +(0,\lenY) node[above] {$f$};

\draw[dashed] ([yshift=\flow]org) to +(\txhigh,0);
\draw[dashed] ([yshift=\fhigh]org) to +(\txhigh,0);

\draw[dashed] ([xshift=\ttx, yshift=\ttylow]org) to ([xshift=\ttx, yshift=\ttyhigh]org)  node[above] {$t_0$};

\node[] at ([xshift=0.5*\ttx,yshift=\ttylow+0.5*(\ttyhigh-\ttylow)]org) {?};
\node[] at ([xshift=\ttx+0.5*(\txhigh-\ttx),yshift=\ttylow+0.5*(\ttyhigh-\ttylow)]org) {?};

\coordinate (l3) at ([xshift=0.5*\lenX, yshift=\lenY]org);

\node[captll, left=of ll1,align=center,text width=1.5cm] {C$^3$ACE};
\node[captll, left=of ll2,align=center,text width=1.5cm] {Spectrum\\ Analyzer};

\node[capt, shape border rotate=270, above=of l1,align=center,text width=1.5cm] {No Load\\ No User};
\node[capt, shape border rotate=270, above=of l2,align=center,text width=1.5cm] {High Load\\ Single User};
\node[capt, shape border rotate=270, above=of l3,align=center,text width=1.5cm] {High Load\\ Many User};

\end{sansmath}
\end{tikzpicture}
	\caption{Simplified illustration of resource occupations when an \UE starts a transmission at time $t_0$ for different cell load scenarios. The granted number of \RBs for that \UE (crosshatched in blue) depends on the cell load and the number of concurrently transferring \UEs. The number of active \UEs is unknown at \UE side and cannot be derived from signal power using a spectrum analyzer. \cccace uncovers this number at \UE side including the particular allocations of each active participant.}
	\label{fig:scenarios}
\end{figure}

As LTE organizes resources in terms of \acfp{RB} in a resource grid with every \RB  consisting of mostly 7 symbols and 12 subcarriers, one could measure the signal power on each \RB and obtain the utilization.
This approach allows the identification of a slightly loaded cell with the expectation of a high data rate for an ongoing transmission as there still remain unused resources (see first column in Fig.~\ref{fig:scenarios}).
However, in case of a fully loaded cell  the number of assigned resources highly depends on the total number of active participants (see second and third column in Fig.~\ref{fig:scenarios}).
With this approach that information stays ambiguous, as illustrated in the first row, because the participants cannot be distinguished from signal power in downlink and uplink utilization might be affected by hidden stations that cannot be perceived by the particular \UE.

In an LTE network, the assignment of the resources for each \UE, designated as \DCI, is carried in the \PDCCH at the beginning of each subframe every \SI{1}{\milli\second}.
Unfortunately, the \PDCCH is designed in a way that a \UE can only decode its own allocations and common allocations containing public data, e.g. system information.
In this paper we present a \acf{C3ACE}, a method to decode and process the entire \PDCCH obtaining the number of active \UEs and the individual resource allocations for both downlink and uplink.
We implemented our approach using an open source \SDR \UE allowing for real-time passive probing.
Furthermore, we show that a joining \UE can reliably estimate its expected data rate based on the disclosed information.

\section{Related Work}
For the evaluation of the current performance of mobile communication system, often active probing mechanisms are applied \cite{Huang}. For doing so, an active network connection and dedicated network load are required in order to evaluate the performance. In contrast to that, the analysis of passive indicators does not need this dedicated network connection. However, the analysis of passive indicators like the \RSRP and \RSRQ provide only a stochastic answer about the correlation between indicator and data rate  \cite{Ide2016a}. Further measurement-based analyses (in public LTE networks) of the correlation between passive performance indicators (e.g., \RSSI) and LTE data rate are presented in \cite{Wu}.
In order to enhance the forecast quality of the data rate, several approaches that analyze the spectrum can be found in the literature.
A performance analysis of 2G mobile networks based on a crowdsourced examination of control traffic by a set of modified smartphones is presented in~\cite{achtzehn2015}.
In \cite{lteye2014} the LTE radio performance is analyzed to discover an inefficient spectrum utilization.
For this purpose, the users are localized within a cell.
However, the localization requires a rotating antenna to create a Synthetic Aperture Radar (SAR).
In addition, this approach does not support a real-time analysis, since a file is periodically written and the results are analyzed offline.
An extension of that system can be found in \cite{piStream2015}.
They provide an adaptive Hypertext Transfer Protocol (HTTP) video streaming over LTE.
In order to decrease the risk of stalling video, the base stations' resource allocation is monitored to get information about the available bandwidth.
Although this is a real-time implementation, it is not possible to capture the number of active users because only the \RB allocation based on the signal power within \RBs is analyzed.
In contrast to that, we will show is this paper that this value is important for a reliable estimation and we can make it available.
The knowledge of the cell occupancy and the number of active users can be leveraged for example to improve the resource efficiency of LTE Machine-Type Communication (MTC) data transmissions in vehicular environments \cite{IdeTVT} that consider the approximation of the mobile connectivity in the \UE.

\section{Resource Partitioning in LTE}\label{ch:scheduling}
In LTE the available radio resources, referred as \RBs, are shared among  active participants and are allocated by the \eNB.
This may happen in different ways, heading for distinct objectives.
The \eNB might share the radio resources in a manner that all active \UEs achieve equal data rates.
Although this approach seems  fair, it has one main disadvantage if few far-distant  \UEs demand for high data rates.
These participants can only reach lower data rates as a consequence of more robust \MCS.
This limits the data rates of nearer \UEs in a significant way compared to their highest achievable rate.
As a result this  would lower the User Experience counteracting marketing promises of high data rates made by the network operators.

Another scheduling strategy is to share the available \RBs equally or with respect to a particular priority class over the demanding \UEs.
In case of similar priorities this leads to high data rates for \UEs near to the \eNB and moderate rates for edge \UEs at low cell loads.
This kind of resource fair scheduling is utilized by the \eNB of our dedicated network which we will describe later in section~\ref{ch:setup}.
In a fully loaded cell with $N$ \UEs of equal priority and the aforementioned scheduling the number of \RBs allocated to each \UE can be estimated by
\begin{eqnarray}
\operatorname{NRB}_{\text{UE}} = \frac{1}{N} \cdot \operatorname{NRB}_{\text{Total}}
\end{eqnarray}
where $\operatorname{NRB}_{\text{Total}}$ is the \eNB's total number of available \RBs per slot.
This results in a data rate
 \begin{eqnarray}
  r_{\text{UE}} =\operatorname{NRB}_{\text{UE}} \cdot  r_0(\operatorname{MCS}_{\text{UE}})
 \end{eqnarray}
 where $ r_0(\operatorname{MCS}_{\text{UE}})$ stands for the achievable  rate through a single \RB at given \MCS.
Based on this an idle \UE can estimate its expected data rate as
\begin{eqnarray}
r_{\text{UE}}^{\text{expected}} = \frac{1}{N+1} \cdot \operatorname{NRB}_{\text{Total}} \cdot r_0(\operatorname{MCS}_{\text{UE}}).\label{eq:datarate}
\end{eqnarray}
While $\operatorname{NRB}_{\text{Total}}$ is known by the \UE and the \MCS can be derived from already existing connectivity indicators like \RSRP and \RSRQ at the \UE side, the number of active participants is only known by the \eNB.
However, we present that $N$ can be determined by analyzing the \eNB's \PDCCH, requiring modifications of the \UE's decoding procedure.

\section{Downlink Control Channel Analysis}
The proposed \cccace approach requires several modifications and additional validation steps within the \UE's control channel decoding procedure.
To understand the upcoming challenges we first introduce the structure of the \PDCCH and the containing data.
Afterwards we briefly present the utilized software and hardware platform in which we implemented a passive probe.
Finally we focus on the validation of the decoded data which is the crucial part of \cccace.

\subsection{Structure of the Channel}\label{ch:channel_structure}
\begin{figure}[b!]  
\centering		  
	\begin{tikzpicture}%
\begin{sansmath}

\tikzstyle{title} = [draw,align=center,font=\sffamily\footnotesize, node distance=3mm]
\tikzstyle{cce} = [draw, inner sep=0pt, minimum width=5mm, minimum height=3mm, font=\tiny, node distance=0]
\tikzstyle{allocated} = [fill=yellow]
\tikzstyle{ccedots} = [inner sep=0pt, minimum width=5mm, minimum height=2mm, font=\sffamily\tiny, node distance=0]

\tikzstyle{ccelabel} = [inner sep=1pt, font=\sffamily\tiny, node distance=2mm, left=2mm]
\tikzstyle{ccelabelr} = [inner sep=1pt, font=\sffamily\tiny, node distance=2mm, right=2mm]
\tikzstyle{wrong} = [red]
\tikzstyle{ccebrace} = [decorate,decoration={brace,mirror,amplitude=2pt,raise=1pt,pre=moveto,pre length=0.3pt,post=moveto,post length=0.3pt}]
\tikzstyle{ccebracer} = [decorate,decoration={brace,amplitude=2pt,raise=1pt,pre=moveto,pre length=0.3pt,post=moveto,post length=0.3pt}]

\node[cce,allocated] (cce0) at (0,0)        {0};
\node[cce, below= of cce0] (cce1) {1};
\node[cce, allocated, below= of cce1] (cce2) {2};
\node[cce, allocated, below= of cce2] (cce3) {3};
\node[cce, below= of cce3] (cce4) {4};
\node[cce, below= of cce4] (cce5) {5};
\node[cce, allocated, below= of cce5] (cce6) {6};
\node[cce, allocated, below= of cce6] (cce7) {7};
\node[ccedots, below= of cce7] (cceK) {\ldots};
\node[cce, below= of cceK] (cceN) {N};

\draw[ccebrace] (cce0.north west) [above left] -- (cce0.south west) node[ccelabel, midway] (exampleNB) {for \texttt{0xFFFF}};
\draw[ccebrace] (cce1.north west) -- (cce1.south west) node[ccelabel, midway] {free};
\draw[ccebrace] (cce2.north west) -- (cce3.south west) node[ccelabel, midway] {for \texttt{0xAAAA}};
\draw[ccebrace] (cce4.north west) -- (cce4.south west) node[ccelabel, midway] {free};
\draw[ccebrace] (cce5.north west) -- (cce5.south west) node[ccelabel, midway] {free};
\draw[ccebrace] (cce6.north west) -- (cce7.south west) node[ccelabel, midway] {for \texttt{0xAAAB}};
\draw[ccebrace] (cceN.north west) -- (cceN.south west) node[ccelabel, midway] {free};

\draw[ccebracer] (cce0.north east) -- (cce0.south east) node[ccelabelr, midway] (example1)        {\texttt{0xFFFF}};
\draw[ccebracer] (cce1.north east) -- (cce1.south east) node[ccelabelr, wrong, midway] {\texttt{0x3754}};
\draw[ccebracer] (cce2.north east) -- (cce2.south east) node[ccelabelr, wrong, midway] {\texttt{0xAD45}};
\draw[ccebracer] (cce3.north east) -- (cce3.south east) node[ccelabelr, wrong, midway] {\texttt{0xBB3E}};
\draw[ccebracer] (cce4.north east) -- (cce4.south east) node[ccelabelr, wrong, midway] {\texttt{0xB23E}};
\draw[ccebracer] (cce5.north east) -- (cce5.south east) node[ccelabelr, wrong, midway] {\texttt{0x98C4}};
\draw[ccebracer] (cce6.north east) -- (cce6.south east) node[ccelabelr, wrong, midway] {\texttt{0x2053}};
\draw[ccebracer] (cce7.north east) -- (cce7.south east) node[ccelabelr, wrong, midway] {\texttt{0x8967}};
\draw[ccebracer] (cceN.north east) -- (cceN.south east) node[ccelabelr, wrong, midway] {\texttt{0xF084}};

\def\aggrshift{1.2cm}
\draw[ccebracer] ([xshift=\aggrshift]cce0.north east) -- ([xshift=\aggrshift]cce1.south east) node[ccelabelr, wrong, midway] (example2)        {\texttt{0x947D}};
\draw[ccebracer] ([xshift=\aggrshift]cce2.north east) -- ([xshift=\aggrshift]cce3.south east) node[ccelabelr, midway] (correctexample) {\texttt{0xAAAA}};
\draw[ccebracer] ([xshift=\aggrshift]cce4.north east) -- ([xshift=\aggrshift]cce5.south east) node[ccelabelr, wrong, midway] (wrongexample) {\texttt{0xB123}};
\draw[ccebracer] ([xshift=\aggrshift]cce6.north east) -- ([xshift=\aggrshift]cce7.south east) node[ccelabelr, midway] {\texttt{0xAAAB}};
\draw[ccebracer] ([xshift=\aggrshift]cceK.east) -- ([xshift=\aggrshift]cceN.south east)node[ccelabelr, wrong, midway] (wrongexample2) {\texttt{0x7769}};

\node[title, above= of cce0, text width=1.4cm, align=center] (cceT) {Control Channel Element (CCE)};
\draw[-stealth] (cceT.south) -- ([yshift=-3mm]cceT.south);

\node[title, left= of cceT, text width=1.5cm, align=center] (Tleft) {\textbf{Allocated} by eNB};
\draw[-stealth] (Tleft.south) -- +(0cm,-0.5cm) -| (exampleNB.north);

\node[title, right= of cceT, text width=2cm, align=center] (Tright) {\textbf{Decoded} User-RNTI at Aggregation 1};
\draw[-stealth] (Tright.south) -- +(0cm,-0.3cm) -| (example1.north);

\node[title, node distance=1mm, right=of Tright, text width=2cm, align=center] (Tright2) {\textbf{Decoded} User-RNTI at Aggregation 2};
\draw[-stealth] (Tright2.south) -- + (0cm, -0.5cm) -| (example2.north);

\node[title, very thick, node distance=1cm, below=of Tright2, text width=1.5cm, align=center] (Correct) {Correct\\ Information\\(Black)};
\draw[-stealth] (Correct.west) -- (correctexample.east);

\node[title, dashed, very thick, wrong, below=of Correct, text width=1.5cm, align=center] (Wrong) {False\\ Information\\(Red)};
\draw[-stealth, wrong] (Wrong.west) -- (wrongexample.east);

\coordinate (pdcchN) at ([xshift=-2cm]cce0.north west);
\coordinate (pdcchS) at ([xshift=-2cm]cceN.south west);
\draw[|stealth-{stealth}|] (pdcchN) -- (pdcchS) node[midway, rotate=90, fill=white] {PDCCH};
\draw[-,dashed] (pdcchN) -- (cce0.north west);
\draw[-,dashed] (pdcchS) -- (cceN.south west);

\end{sansmath}
\end{tikzpicture}
	\caption{Structure of the \acf{PDCCH}, consisting of numerous \acp{CCE}. The \eNB may place control information at different aggregation levels and leave some \acp{CCE} unused forcing the receiver to scan the entire \ac{PDCCH} at all aggregation levels. Without the knowledge of any valid \RNTIs this results in a huge number of invalid information (red).}
	\label{fig:pdcch}
\end{figure}
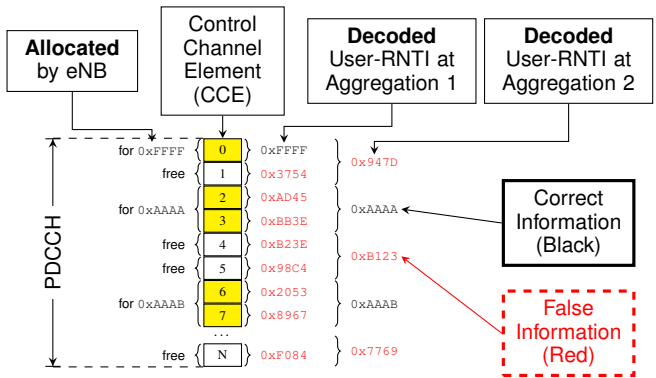

The \PDCCH bears information about the allocation of uplink and downlink resources for particular \UEs as well as for common system information about the cell.
It is transmitted every \SI{1}{\milli\second} allocating the \RBs of the following LTE subframe.
Fig.~\ref{fig:pdcch} illustrates the structure of the channel which is divided in a set of \CCEs containing the actual resource allocations.
To obtain a higher degree of robustness against transmission errors the \eNB may aggregate 1, 2, 4, or 8 consecutive \CCEs for one chunk of information using a higher code rate.
Furthermore, each chunk contains a \CRC allowing an integrity check for the received data.
This \CRC is additionally scrambled with the \RNTI of the \UE to which it is addressed to.

A usual \UE obtains its allocation as follows:
After attaching to the \eNB and receiving its specific \RNTI over an encrypted channel, the \UE continuously decodes single \CCEs and compares the resulting checksum to its assigned \RNTI.
In case of no match, or if the \UE expects further information, it repeats this procedure decoding two consecutive \CCEs at once.
This is also done for 4 and 8 aggregated \CCEs.

Without the knowledge of valid \RNTIs a receiver cannot validate the integrity of the decoded chunk.
Furthermore, every decoding process at all aggregation levels results in a possibly random sequence of bits as a consequence of noise or a mismatching aggregation (cf. red entries in Fig.~\ref{fig:pdcch}).

In addition, the actual information carried by \CCEs can consist of data structures with different lengths as a result of different \DCI formats~\cite{36.213}.
These formats serve specific purposes, e.g., compact scheduling or multi-user MIMO scheduling.
They are normalized by the \eNB using rate adaption to fit into whole \CCEs thus making them 	indistinguishable on the channel and demanding the \UE for a \emph{trial and error} decoding for all possible formats.

\subsection{Implementation Platform}
The proposed passive probe is based on \textsc{OpenAirInterface}'s (\acs{OAI}) \cite{OAI} application \textsc{lte-softmodem},
an open source \ac{SDR} implementation of \eNB and \UE.
It supports several radio front ends, including the Ettus USRP B210 platform, which we chose.

In order to build our probe, we modified the \UE to keep synchronized to a specified \eNB without making any active access (e.g. attach requests) and to continuously decode the whole \PDCCH.
The results of the analysis are logged to a file and printed on the screen in real time.

\subsection{DCI validation}\label{ch:DCI_validation}
As each \DCI's checksum is scrambled with the \RNTI to which it is addressed, the probe cannot validate the integrity of the \DCI without the knowledge of valid \RNTIs (cf. Fig~\ref{fig:pdcch}).
Furthermore, the probe will receive a random bit sequence when decoding unused \CCEs, mismatching aggregation levels, and incorrect \DCI formats as explained in section~\ref{ch:channel_structure}.
Hence, the blind analysis %
will result in a high number of possible \DCI with the majority containing invalid information.

In order to remove these invalid \DCI we applied a two-step filtering: First we removed impossible \DCI, as the \eNB may place a \DCI only in a specific subset of \acp{CCE} depending on the current subframe index ($0\ldots{}9$) and the \RNTI it is addressed to \cite{36.213}.
This feature is designed to reduce the search complexity for a normal \UE by approximately a fifth in a \SI{10}{\mega\hertz} cell and a tenth in a \SI{20}{\mega\hertz} cell.
In our approach we use it to disqualify $80\%$ at \SI{10}{\mega\hertz} and  $90\%$ at \SI{20}{\mega\hertz} of received \DCI without dropping any valid information.

In the second step we removed all \DCI with a high likelihood of containing false information.
At this step we evaluated two different approaches of rating the likelihood.
The first method was inspired by \cite{lteye2014} where each decoded information is being convolutionally re-encoded and compared to the initially received encoded sequence.
If the sequences differ by more than few bits, the information is rated as false.
Our experiments showed a reliable classification with a threshold of two bit errors per \CCE under optimal radio conditions, placing the receiver next to our dedicated \eNB. 
Under normal conditions, which is the synchronization to a public \eNB out of our office, a minimum threshold of about twelve bit errors per \CCE was required to classify any valid \DCI\footnote{We evaluated this by exclusively decoding periodic system information having a dedicated \RNTI of \hex{FFFF}.}.
With this threshold we observed at least one double-allocation collision of a resource block per subframe, which results in processing false \DCI.

In the second approach we first determine valid \RNTIs out of the entire value range of possible candidates.
For this we assume true \RNTIs appear more frequently in the recent period of time while false \RNTIs appear equally spread over the value range.
The distinction is made by maintaining a histogram of the last $n$ occurred \RNTIs and classifying as true those \RNTIs which exceed a threshold value $k$ in the histogram.

\begin{figure}[tb]  	
\centering		  
    \begingroup%
    \StrCount{figures/histogram}{/}[\matches]%
    \StrBefore[\matches]{figures/histogram}{/}[\datapath]%
    \begin{tikzpicture}[spy using outlines={circle, magnification=4, connect spies}]
\begin{sansmath}
\begin{axis}[
		ybar,
		bar width=0.1pt,
		width= 235.0pt,			%
		height=5cm,
		xmin=0, xmax=66000,
		ymin=0, ymax=70,
		xticklabel style={font=\tt},
		xtick={0,16384,32768,49152,65536},
		xticklabels={0x0,0x4000,0x8000,0xC000,0xFFFF},
		scaled x ticks=false,		%
		extra y ticks={8},
		extra y tick style={grid=major, ticklabel pos=lower},
		extra y tick labels={$k$},
		title={\textsc{RNTI Histogram}},
		xlabel={Observed RNTI Value by C$^3$ACE}, 
		ylabel={Occurences within Window $\left[\frac{1}{T}\right]$},
		ylabel near ticks,
		grid=both,
        ]

\addplot[color=blue, fill=blue] file {\datapath/by50-500ms.txt};
\draw[-,red,dashed] (axis cs:0,8) to (axis cs:66000,8);

\coordinate (high) at (axis cs: 53000,50);
\coordinate (low) at (axis cs: 51000,15);
\coordinate (threshold) at (axis cs: 100,8);
\coordinate (arbitrary) at (axis cs: 25000,5);

\end{axis}

\tikzstyle{mylabel} = [draw,fill=white]

\draw[stealth-] (high) to (5.0cm, 3cm)  node[mylabel, left] {Highly Active UE} ;
\draw[stealth-] (low) to (4.8cm, 2.3cm)  node[mylabel, left] {Lowly Active UEs} ;
\draw[stealth-] (threshold) to (1cm, 1.3cm)  node[mylabel, fill, above] {Threshold} ;
\draw[stealth-] (arbitrary) to (3cm, 0.7cm)  node[mylabel, fill, above] (f) {False Occurences} ;
\draw[stealth-] ([xshift=0.5cm]arbitrary) to (f.south);
\draw[stealth-] ([xshift=1cm]arbitrary) to (f.south);

\end{sansmath}
\end{tikzpicture}%
    \endgroup%

	\caption{Snapshot of an \RNTI histogram while analyzing the load on a public cell. Active \UEs clearly appear as \RNTIs with a high number of recent occurrences. False \RNTIs are uniformly spread over the value range allowing for a threshold-based filtering.}
	\label{fig:histogram}
	\vspace{-5pt}
\end{figure}
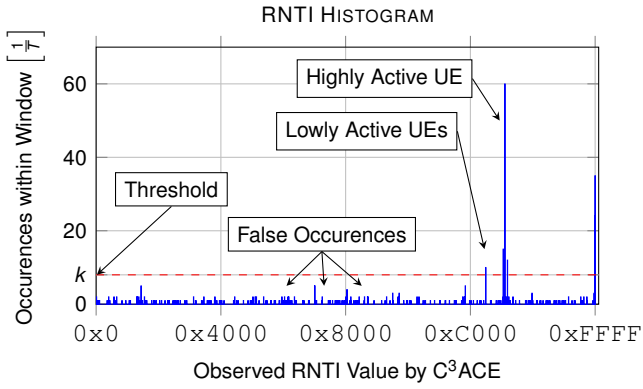

A snapshot of such a histogram is shown in Fig.~\ref{fig:histogram} during the load analysis of a public cell.
It clearly shows certain \RNTIs appear more often as a consequence of the dedicated transmissions while arbitrary \RNTIs are uniformly distributed, turning up as noise.
Moreover, the figure clarifies that the delay for detecting a true positive and the probability for a false positive depend on the chosen threshold value $k$: A too low threshold will result in a quick true positive detection but also in a high amount of false positive classifications and vice versa.

\begin{figure}[tb]  	
\centering		  
    \begingroup%
    \StrCount{figures/invprop}{/}[\matches]%
    \StrBefore[\matches]{figures/invprop}{/}[\datapath]%
    \begin{tikzpicture}[spy using outlines={circle, magnification=4, connect spies}]
\begin{sansmath}
\begin{axis}[
		width= 235.0pt,			%
		height=5cm,
		xmin=0, xmax=20,
		ymin=1e-7, ymax=1.0,
		ymode=log,
		scaled x ticks=false,		%
		ytick={1e-6, 1e-3, 1e0},
		extra x ticks={8},
		extra x tick style={grid=major},
		extra y tick style={grid=major, ticklabel pos=lower},
		title={\textsc{Probability for a False Decetion per Subframe}},
		xlabel={Threshhold Value $k$ in $\left[\frac{1}{T}\right]$}, 
		ylabel={Probability},
		ylabel near ticks,
		grid=both,
		legend cell align=left,
		legend pos=north east,
        ]
        
\addlegendimage{empty legend}
\addlegendentry{\hspace{-0.6cm}Histogram Length}
\addplot[color=blue, mark=x] file {\datapath/invprop100.txt};
\addlegendentry{$T=\SI{100}{\milli\second}$}
\addplot[color=green, mark=triangle] file {\datapath/invprop200.txt};
\addlegendentry{$T=\SI{200}{\milli\second}$}
\addplot[color=red, mark=o] file {\datapath/invprop500.txt};
\addlegendentry{$T=\SI{500}{\milli\second}$}
\addplot[color=black, mark=square] file {\datapath/invprop1000.txt};
\addlegendentry{$T=\SI{1000}{\milli\second}$}

\addplot[color=red, mark=o] file {\datapath/invprop500.txt};	%

\draw[-,red, dashed] (axis cs: 8,1e-7) -- (axis cs: 8,1);
\draw[stealth-] (axis cs:8.1,5.55002617e-07) -- (4cm, -2.8cm) node[draw, fill=white,right] {$5.6\cdot 10^{-7}$};

\end{axis}

\tikzstyle{mylabel} = [draw,fill=white]

\end{sansmath}
\end{tikzpicture}%
    \endgroup%

	\caption{Probability for a classification error during the analysis of a subframe as function of the chosen threshold value $k$ for various histogram lengths.
In case of \SI{500}{\milli\second} a threshold value $k=8$ leads to an error probability of $5.6\cdot 10^{-7}$ for a false detection of any arbitrary \RNTI per subframe.}
	\label{fig:probability}
	\vspace{-5pt}
\end{figure}
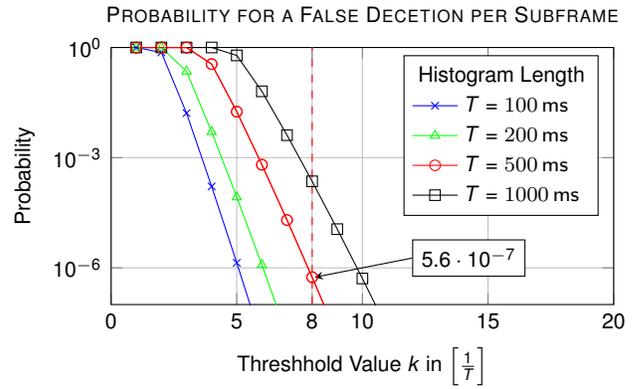

In order to determine the probability for a false positive classification we assume a uniform distribution of false \RNTIs.
The probability for a particular $\text{\RNTI} = x$ to result from the decoding of an empty or otherwise used \CCE is
\begin{eqnarray}
P(\text{\RNTI} = x) = \frac{1}{2^{16}} = p
\end{eqnarray}
as a fact of $2^{16}$ possible \RNTIs. Within a histogram over $n$ recent \RNTIs the probability for a single \RNTI value to appear less or equally than $k$ times is
\begin{eqnarray}
P(Z \leq k) = \sum_{i=0}^{k} \binom{n}{i} \cdot p^i \cdot (1-p)^{n-i}\label{eq:bernoulli}
\end{eqnarray}
which follows from the Bernoulli distribution of $n$ coin tosses.
As the calculation of binomial coefficients quickly tends to deal with very large numbers, the probability can be approximated by the Poisson limit theorem \cite{Papoulis} for large $n$ and small $p$ as follows:
\begin{eqnarray}
\binom{n}{k} \cdot p^k \cdot (1-p)^{n-k} \simeq e^{-np} \frac{(np)^k}{k!}\label{eq:poisson}
\end{eqnarray}
which leads to
\begin{eqnarray}
P(Z \leq k) \simeq e^{-np} \sum_{i=0}^{k} \frac{(np)^i}{i!}\label{eq:approx}.
\end{eqnarray}
Finally, the probability for all $2^{16}$ possible \RNTI values to occur $k$ or less times within a histogram is
\begin{eqnarray}
P_{\text{all}}(Z \leq k) = P(Z \leq k)^{(2^{16})}.
\end{eqnarray}
In order to identify a reliable threshold value $k$ we calculated $P_{\text{all}}(Z \leq k)$ for different histogram lengths as shown in Fig.~\ref{fig:probability}.
The corresponding time interval $t$ in seconds covered by the respective histogram is
\begin{eqnarray}
t = \frac{1}{1000} \cdot N_{\text{\RNTIs\_per\_subframe}} \cdot p_{\text{Filter1}}
\end{eqnarray}
where $N_{\text{\RNTIs\_per\_subframe}}$ is the number of upcoming \RNTI candidates per subframe and $p_{\text{Filter1}}$ the filtering rate by the first \RNTI-filtering stage described at the beginning of section~\ref{ch:DCI_validation}.
In case of a \SI{10}{\mega\hertz} cell and scanning for DCI formats 0, 1, and 1A \cite{36.212} these values are typically
\begin{eqnarray}
N_{\text{\RNTIs\_per\_subframe}} = 164, \qquad p_{\text{Filter1}} = \frac{1}{5},
\end{eqnarray}
which we used in Fig.~\ref{fig:probability}.
It shows that the probability for an error-free subframe, without any false positive \RNTI, quickly converges to $1$.
A threshold of $k=8$ with a histogram length of $t=\SI{500}{\milli\second}$ leads to a probability of $5.6\cdot 10^{-7}$ for a false positive detection of any \RNTI per subframe.
Consequently any new upcoming \UEs will be detected with a delay of $k=8$ occurrences which is \SI{8}{\milli\second} if its \RNTI appears once in eight consecutive subframes.
While keeping the probability for an error at the same level, the latency might be lowered by reducing the histogram length which consequently lowers the required threshold.
A shorter histogram on the other hand leads to a lower recognition rate of sporadically active \UEs as their frequency of occurrence may not suffice to exceed the threshold.
For our measurements we chose $t=\SI{500}{\milli\second}$ and $k=8$. %

\section{Measurements}\label{ch:setup}
In order to show the benefits of \cccace, we performed data rate estimations based on the discovered number of active \UEs in a cell.

\subsection{Measurement Setup}

\begin{table}[!t]
\renewcommand{\arraystretch}{1.3}
\caption{Properties of the Dedicated LTE Test Network}
\label{tab:properties}
\centering
\begin{tabular}{lr}
\hline
Frequency Band & \SI{2.6}{\giga\hertz} (Band 7)\\
EIRP & \SI{158}{\milli\watt} (\SI{22}{\dBm})\\	%
Bandwidth & \SI{10}{\mega\hertz} (50 \RBs)\\
Duplex Mode & Frequency Division Duplex\\
\hline
\end{tabular}
\vspace{-5pt}
\end{table}

\begin{figure}[tb]  	
\centering		  
	\includegraphics[height=4cm]{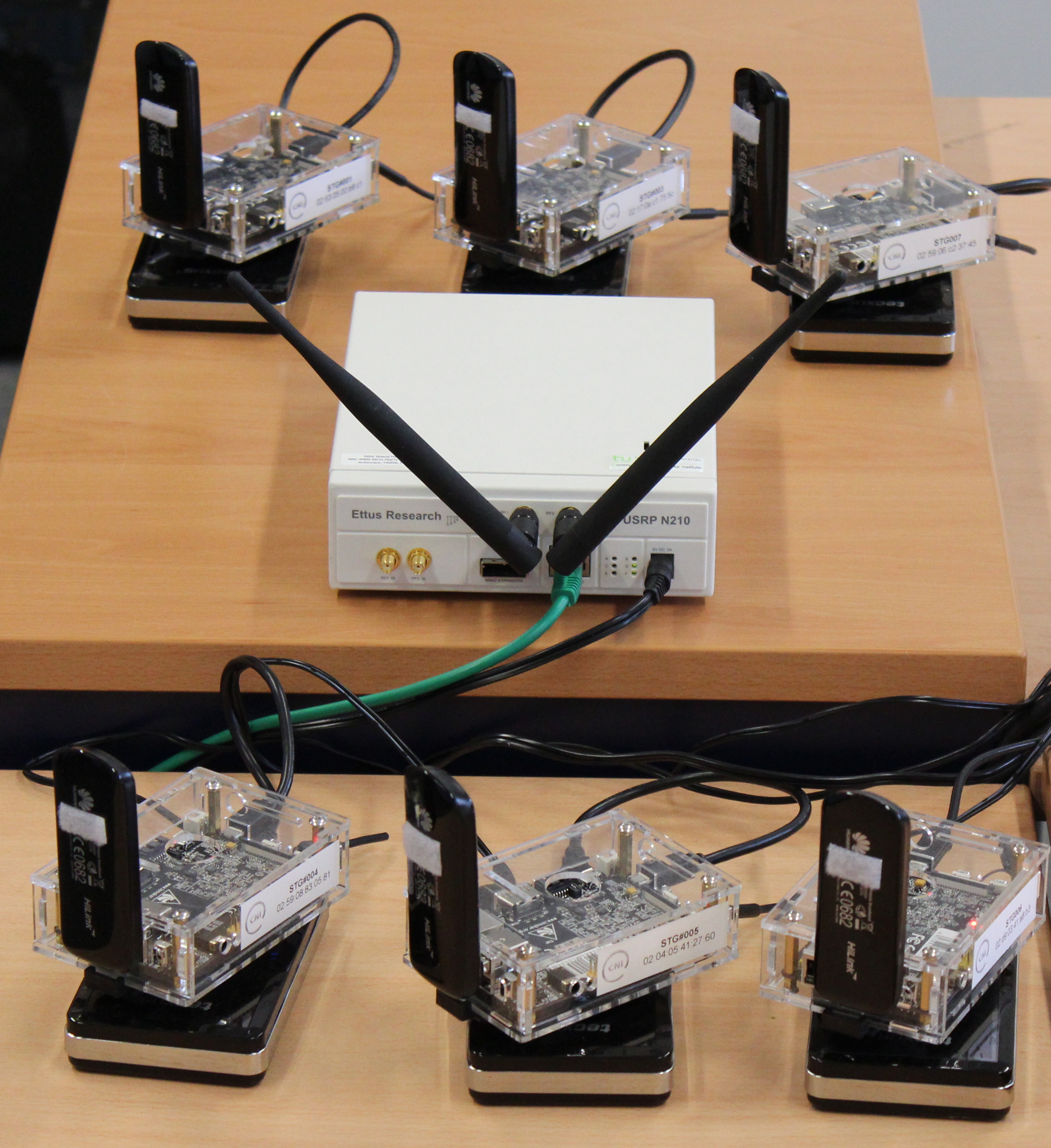}
	\qquad
	\includegraphics[height=4cm]{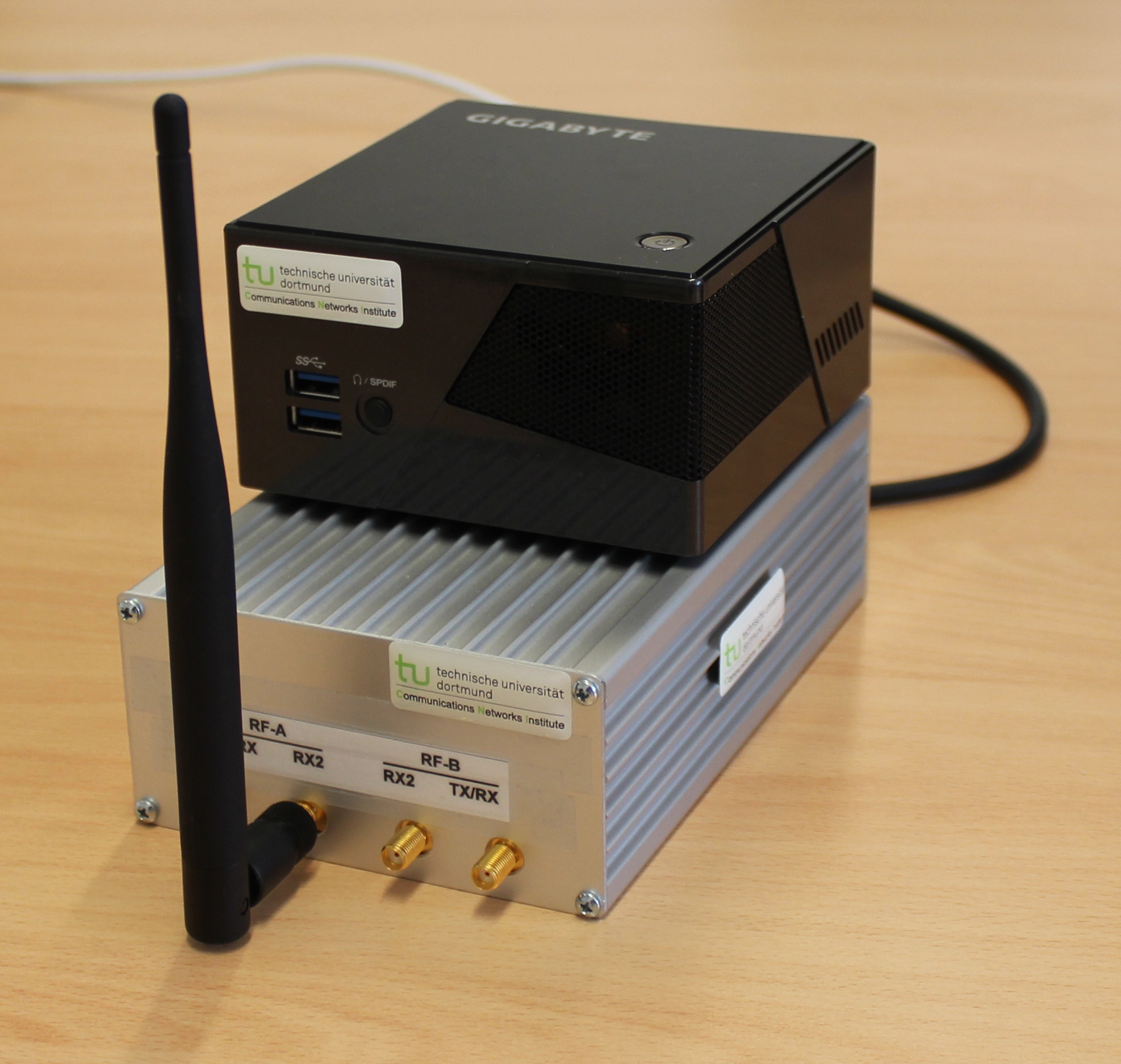}
	\caption{Left: Measurement setup of the \emph{near} scenario using an USRP N210 as \eNB and six  \acp{STG} including the \DUT. Right: LTE passive probe consisting of an USRP B210 and a compact PC executing the modified \OAI{}-software.}
	\label{fig:setup-photo}
\end{figure}

For our measurements, we used a dedicated LTE network with a configuration as listed in Tab.~\ref{tab:properties}.
Due to our research license in the chosen frequency band no other interfering networks were in range.
Fig.~\ref{fig:setup-photo} shows a photo of the setup consisting of an \eNB, several \UEs denoted as \STGs~\cite{Kaulbars2015} including a \DUT, and the passive probe which continuously analyses the \PDCCH.
The \STGs, consisting of an embedded computer and an LTE modem, are used to induce selected traffic load into the cell.
As the implementation of the \textsc{OpenAirInterface}'s \UE was not fully completed during this work, we switched to a dedicated \DUT which is synchronized to the passive probe.
The \eNB is a commercial \SDR implementation by Amarisoft using an USRP N210 as radio head.

\begin{figure}[tb]  	
\centering		  
	\includegraphics[scale=1]{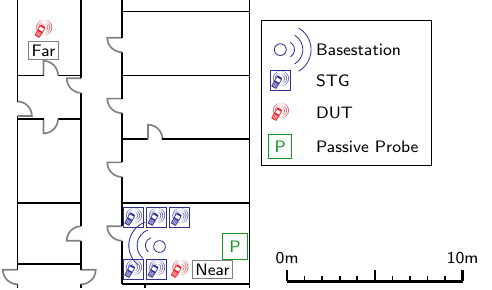}
	\caption{Overview of the measurement setups for \emph{near} and \emph{far} scenario in our office environment. The corresponding radio conditions are listed in Tab.~\ref{tab:positionproperties}.}
	\label{fig:floor}
	\vspace{-5pt}
\end{figure}

For a measurement we instructed up to five \STGs to transmit as much data as possible in one direction utilizing \iperf~\cite{iperf}.
Then the passive probe logged the detected number of currently active users to a file emulating an \UE evaluating its connectivity for its upcoming transmission.
Afterwards the \DUT initiated an additional \iperf{}-transmission logging the actually achieved data rate.
We performed this measurement for a varying number of active \STGs, and for poor and good radio conditions by placing the \DUT far away or next to the \eNB as shown in Fig~\ref{fig:floor}.
The corresponding connectivity indicators of the \DUT are listed in Tab.~\ref{tab:positionproperties}.
Each single measurement was repeated 50 times.

\begin{table}[!t]
\renewcommand{\arraystretch}{1.3}
\caption{Radio Conditions Measured by the DUT}
\label{tab:positionproperties}
\centering
\begin{tabular}{lrr}
\bfseries Indicator & \bfseries Near & \bfseries Far\\
\hline
\RSRP & \SI{-86}{\dBm} & \SI{-129}{\dBm}\\
\RSSI & \SI{-65}{\dBm} & \SI{-100}{\dBm}\\
\RSRQ & \num{-3} & \num{-15}\\
\hline
\end{tabular}
\end{table}

\subsection{Measurement Results}
\begin{figure}[tb]  	
\centering		  
    \begingroup%
    \StrCount{figures/boxplot-rate}{/}[\matches]%
    \StrBefore[\matches]{figures/boxplot-rate}{/}[\datapath]%
    \begin{tikzpicture}%
\begin{sansmath}
\begin{axis} [
		title={\textsc{Near} (RSRP = -86dBm)},
		width=5cm,
		height=5cm,
		xmin=-0.5, xmax=5.5,
		ymin=0, ymax=11,
		xtick={0,...,5},
		minor y tick num=1,
		ylabel={Achieved Uplink Rate $\left[\si{\mega\bit\per\second}\right]$},
		grid=none,
		ymajorgrids,
		yminorgrids,
		alias=AX,
        ]
            
    \addplot [forget plot,thick,color=red,box plot median] table {\datapath/optimal-UL.mat};
    \addplot [forget plot,box plot box] table {\datapath/optimal-UL.mat};
    \addplot [forget plot,box plot top whisker] table {\datapath/optimal-UL.mat};
    \addplot [forget plot,box plot bottom whisker] table {\datapath/optimal-UL.mat};
    \addplot [forget plot,color=black,only marks, mark=+] table {\datapath/optimal-UL-outliers.mat};
    
    \addplot [dashed] table {\datapath/optimal-UL-theory.mat};
    \addlegendentry{\tiny $\frac{\SI{10}{\mega\bit\per\second}}{N+1}$}
    
\end{axis}

\begin{axis} [
		at=(AX.south east), anchor=south west, xshift=3mm,
		title={\textsc{Far} (RSRP = -129dBm)},
		width=5cm,
		height=5cm,
		xmin=-0.5, xmax=5.5,
		ymin=0, ymax=11,
		xtick={0,...,5},
		minor y tick num=1,
		yticklabel pos=right,
		grid=none,
		ymajorgrids,
		yminorgrids,
		alias=myAxis,
        ]

    \addplot [forget plot,thick,color=red,box plot median] table {\datapath/DUT-far-UL.mat};
    \addplot [forget plot,box plot box] table {\datapath/DUT-far-UL.mat};
    \addplot [forget plot,box plot top whisker] table {\datapath/DUT-far-UL.mat};
    \addplot [forget plot,box plot bottom whisker] table {\datapath/DUT-far-UL.mat};
    \addplot [forget plot,color=black,only marks, mark=+] table {\datapath/DUT-far-UL-outliers.mat};

    \addplot [dashed] table {\datapath/DUT-far-UL-theory.mat};
    \addlegendentry{\tiny $\frac{\SI{7.7}{\mega\bit\per\second}}{N+1}$}  
    
\end{axis}

\node[node distance=4mm, below=of myAxis.south west] (myxlabel)  {Number of active UEs detected by C$^3$ACE};

\end{sansmath}
\end{tikzpicture}%
    \endgroup%

	\caption{Achieved uplink data rate of the \DUT depending on the detected number of active participants by the passive probe in the moment just before the \DUT initiated the transmission. The dashed line shows the expected data rate according to Eq.~\ref{eq:datarate}.}
	\label{fig:results-rate}
	\vspace{-5pt}
\end{figure}
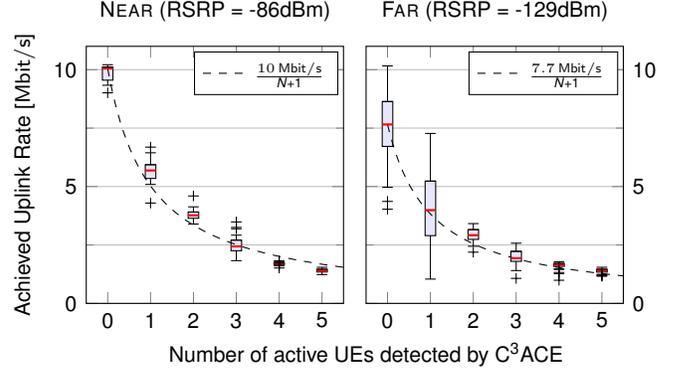
The achieved uplink data rates of the \DUT depending on the detected number of active users by the proposed passive probe are shown in Fig.~\ref{fig:results-rate} for both, good and poor radio conditions, titled as near and far.
In addition, the estimated data rate according to a resource fair scheduling (cf. Eq.~\ref{eq:datarate}) is plotted as dashed graph.
Due to space limitations and greater clarity, we omitted the results for downlink data rates as they showed similar behaviors.

\begin{figure}[tb]  	
\centering		  
    \begingroup%
    \StrCount{figures/boxplot-error}{/}[\matches]%
    \StrBefore[\matches]{figures/boxplot-error}{/}[\datapath]%
    \begin{tikzpicture}%

\begin{axis} [
		title={\textsc{Near} (RSRP = -86dBm)},
		width=5cm,
		height=5cm,
		xmin=-0.5, xmax=5.5,
		ymin=-4, ymax=4,
		xtick={0,...,5},
		minor y tick num=1,
		ylabel={Estimation Error $\left[\si{\mega\bit\per\second}\right]$},
		grid=none,
		ymajorgrids,
		yminorgrids,
		alias=BX,
        ]
        
    \addplot [thick,color=red,box plot median] table {\datapath/optimal-UL-error.mat};
    \addplot [box plot box] table {\datapath/optimal-UL-error.mat};
    \addplot [box plot top whisker] table {\datapath/optimal-UL-error.mat};
    \addplot [box plot bottom whisker] table {\datapath/optimal-UL-error.mat};
    \addplot [color=black,only marks, mark=+] table {\datapath/optimal-UL-error-outliers.mat};
    
\end{axis}
\begin{axis} [
		at=(BX.south east), anchor=south west, xshift=3mm,
		title={\textsc{Far} (RSRP = -129dBm)},
		width=5cm,
		height=5cm,
		xmin=-0.5, xmax=5.5,
		ymin=-4, ymax=4,
		xtick={0,...,5},
		minor y tick num=1,
		yticklabel pos=right,
		grid=none,
		ymajorgrids,
		yminorgrids,
		alias=myAxis,
        ]
        
    \addplot [thick,color=red,box plot median] table {\datapath/DUT-far-UL-error.mat};
    \addplot [box plot box] table {\datapath/DUT-far-UL-error.mat};
    \addplot [box plot top whisker] table {\datapath/DUT-far-UL-error.mat};
    \addplot [box plot bottom whisker] table {\datapath/DUT-far-UL-error.mat};
    \addplot [color=black,only marks, mark=+] table {\datapath/DUT-far-UL-error-outliers.mat};
    
\end{axis}

\node[node distance=4mm, below=of myAxis.south west] (myxlabel)  {Number of active UEs detected by C$^3$ACE};

\end{tikzpicture}%
    \endgroup%

	\caption{Corresponding estimation error from expected data rate (c.f. Fig.~\ref{fig:results-rate}). It shows that the expected data rate for an \UE can be derived from the now available number of active \UEs with high accuracy. Poor radio conditions result in a higher error variance.}
	\label{fig:results-error}
	\vspace{-5pt}
\end{figure}
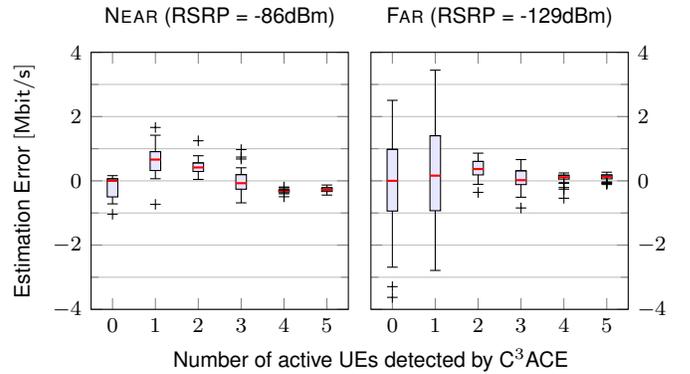
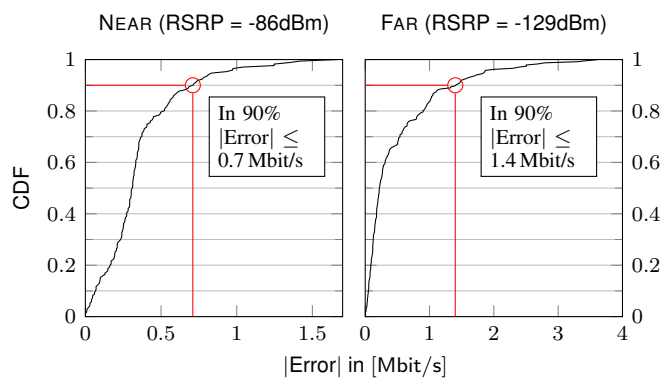
\begin{figure}[tb]  	
\centering		  
    \begingroup%
    \StrCount{figures/boxplot-cdf}{/}[\matches]%
    \StrBefore[\matches]{figures/boxplot-cdf}{/}[\datapath]%
    \begin{tikzpicture}%

\begin{axis} [
		title={\textsc{Near} (RSRP = -86dBm)},
		width=5cm,
		height=5cm,
		xmin=0, xmax=1.7,
		ymin=0, ymax=1,
		minor y tick num=1,
		ylabel={CDF},
		grid=none,
		ymajorgrids,
		yminorgrids,
		alias=CX,
        ]
        
    \addplot [] table {\datapath/optimal-UL-error-cdf.mat};

	\node[red,draw,circle,minimum width=2mm, inner sep=0] (mymark) at (axis cs:0.71,0.9) {};
	\draw[red] (mymark) -- (axis cs:0.71,0);
	\draw[red] (mymark) -- (axis cs:0,0.9); 
	
	\node[draw,fill=white,font=\footnotesize,text width=1.3cm] at (axis cs: 1.2, 0.7) {In 90\% $\left|\text{Error}\right|$ $\leq$ \SI{0.7}{\mega\bit\per\second}};

\end{axis}
\begin{axis} [
		at=(CX.south east), anchor=south west, xshift=3mm,
		title={\textsc{Far} (RSRP = -129dBm)},
		width=5cm,
		height=5cm,
		xmin=0, xmax=4,
		ymin=0, ymax=1,
		minor y tick num=1,
		yticklabel pos=right,
		grid=none,
		ymajorgrids,
		yminorgrids,
		alias=myAxis,
        ]
        
    \addplot [] table {\datapath/DUT-far-UL-error-cdf.mat};

	\node[red,draw,circle,minimum width=2mm, inner sep=0] (mymark) at (axis cs:1.4,0.9) {};
	\draw[red] (mymark) -- (axis cs:1.4,0);
	\draw[red] (mymark) -- (axis cs:0,0.9); 
	
	\node[draw,fill=white,font=\footnotesize,text width=1.3cm] at (axis cs: 2.7, 0.7) {In 90\% $\left|\text{Error}\right|$ $\leq$ \SI{1.4}{\mega\bit\per\second}};
    
\end{axis}

\node[node distance=4mm, below=of myAxis.south west] (myxlabel)  {$\left|\text{Error}\right|$ in $\left[\si{\mega\bit\per\second}\right]$};

\end{tikzpicture}%
    \endgroup%

	\caption{Cumulative distribution function of the error shown in Fig.~\ref{fig:results-error}. At good radio conditions 90\% of estimations have an error smaller or equal than \SI{0.7}{\mega\bit\per\second}.}
	\label{fig:results-cdf}
	\vspace{-5pt}
\end{figure}

To highlight the estimation error, the difference between estimated and achieved data rate is plotted in Fig.~\ref{fig:results-error}.
It shows that the error never exceeds \SI{1.7}{\mega\bit\per\second} during our measurements at good radio conditions.
Evaluating the cumulative distribution function of the estimation error in Fig.~\ref{fig:results-cdf} shows that 90\% of estimations based on \cccace have a lower or equal error than \SI{0.7}{\mega\bit\per\second} at good radio conditions.
At greater distances the connection suffers of higher packet loss rates and re-transmissions as a result of the poorer radio link and lead to a larger error deviation.
As each participant uses a \TCP stream for its transmission, the particular \TCP congestion control significantly reduces its data rate in case of packet losses.
This spares other \UEs additional resources and results in an oscillating data rate.
With a larger number of concurring streams, this effect decreases.
Still, even in those cases, the estimation stays accurate in average (cf. median in the right chart of Fig.~\ref{fig:results-error}) by having an error of less or equal than \SI{1.4}{\mega\bit\per\second} in 90\% of estimations.

\section{Conclusion}
In this paper we proposed a \acf{C3ACE} technique that enables an \UE for a passive determination of the number of active participants within an LTE cell, although an LTE network is not designated to provide this information to an \UE.
Based on this significant information, and with the knowledge of the \eNB's scheduling algorithm, the \UE can estimate at any moment its expected data rate for an upcoming down- and uplink transmission without the need of active probing.
We implemented \cccace into an open source \SDR \UE turning it into a passive probe which is capable of decoding the entire \acf{PDCCH} in real-time.
This uncovers the total resource allocations broken down by individual participants.
The crux of the matter is a reliable validation of blind decoded control information without the knowledge of any assigned \UE-identifier (\RNTI) which prevents a checksum verification.
We solved this by a histogram-based validation technique, which allows for an adjustable reliability.
This approach has a low computational complexity and therefore allows for an efficient integration into future commercial \UE.

Afterwards, we performed measurements to evaluate the quality of data rate estimation based on the detected number of active \UEs within the cell.
The promising results depict an estimation error of less or equal than \SI{0.7}{\mega\bit\per\second} in 90\% of cases at good radio conditions within a \SI{10}{\mega\hertz} cell.

Furthermore, the results bring forward the argument that future mobile networks should implement an indicator for the maximum number of currently assignable resources to an \UE especially for energy and resource constrained Cyber-Physical Systems.
From this, together with the individual radio conditions and the capabilities of the \UE (i.e. multiple antennas), the mobile system could accurately evaluate its expected data rate and decide whether to offload data to the cloud or to perform the necessary computations locally.

In future works we will extend our data rate estimation to be independent of the \eNB's scheduler in consideration of the individual resource demands by active \UEs.

\section*{Acknowledgment}
Part of the work on this paper has been supported by Deutsche Forschungsgemeinschaft
(DFG) within the Collaborative Research Center SFB 876 ``Providing Information by Resource-Constrained
Analysis'', project A4.

\renewcommand{\baselinestretch}{1.04}

\bibliographystyle{IEEEtran}
\bibliography{Bibliography}
\end{document}